\documentstyle[aps,twocolumn]{revtex}
\def\ar{Ar_{13}}
\newcommand{\sbr}{\left( {\sigma\over r_{ij}} \right)}
\newcommand{\be}{\begin{equation}}
\newcommand{\ee}{\end{equation}}
\begin{document}
\draft

\title{\bf\normalsize RELAXATION OF COLLECTIVE EXCITATIONS IN LJ-13 CLUSTER}
\author{Umesh A. Salian\thanks{umesh,snb@iopb.ernet.in }
 S. N. Behera$^*$ and V. S. Ramamurthy}
\address{Institute of Physics, Sachivalaya Marg, Bhubaneswar - 751 005,
INDIA }
\maketitle
\begin{abstract}
We have performed classical molecular dynamics simulation of $Ar_{13}$ cluster
to study the behavior of collective excitations. In the solid ``phase''
of the cluster, the collective oscillation of the monopole mode can be
well fitted to a damped harmonic oscillator. The parameters of the
equivalent damped harmonic oscillator-- the damping coefficient,
spring constant, time period of oscillation and the mass of the oscillator
-- all show a sharp change in behavior at a kinetic temperature of about
$7.0^oK$. This marks yet another characteristic temperature of the system,
a temperature $T_s$ below which collective excitations are very stable, 
and at higher temperatures the single particle excitations cause the 
damping of the collective oscillations. We argue that so long as the cluster
remains confined within the global potential energy minimum the collective
excitations do not decay; and once the cluster comes out of this well,
the local potential energy minima pockets act as single particle excitation
channels in destroying the collective motion. The effect is manifest in
almost all the physical observables of the cluster.
\end{abstract}

\narrowtext
\newpage
\section{Introduction}

  The study of collective excitations
  in finite systems originated in Nuclear Physics and subsequently 
  were carried out in free atoms and metal clusters as well.
  It is understood that in case of atomic nuclei, the coherent 
  motion of shell nucleons driven
  by a sufficiently confining short-range interaction potential gives 
  rise to  collective resonances in specific modes, and these 
  long-lived, large amplitude oscillations are termed ``giant
  resonances''. While single particle excitations are adequately
  described by Hartree-Fock approximation, its time-dependent
  extension accounts for the collective behavior of the 
  constituent fermions \cite{Wang,Cassing}, and the
  damping of the giant resonances is explained by the
  non-perturbative treatment of the residual collisions of
  quasi particles \cite{Blasio}. In case of atoms and metal
  clusters the valence elctrons execute the resonance motion.
  Though long-ranged, the screening makes the Coulomb
  potential effectively short-ranged, enabling the local
  confinement of the electrons that is necessary for the
  resonance to occur \cite{Brech}.

  In this article we present some results of the study of collective
  motions of yet another finite system -- a 13 particle inert gas atom cluster.
  Being a classical, non-fermionic and very weakly interacting system,
  the situation is quite different here. In nuclei and metal clusters,
  the collective excitations are essentially given by the poles of
  the Green's function corresponding to the response of
  the charge density fluctuations, and are amenable to detection by the
  electro-magnetic probe, whereas the constituents of the system
  under present investigation are charge-neutral point particles,
  so there is no possibility of charge fluctuations.
  Nevertheless, the particles do exhibit coherent motion
  owing to the presence of interaction potential between them,
  and these are simply the normal modes of the mass density
  fluctuations. In section II we will describe how 
  the normal mode components of the cluster can be projected out of the 
  complex dynamics that it undergoes.

 It is well know that molecular dynamics simulations of $\ar$ cluster 
 display solid-liquid phase transitions. The cluster has a near-rigid
 regular icosahedron structure in the solid-like ``phase'',
 the individual atoms executing vibrational motions about their
 mean positions, whereas in the liquid-like phase the 
 cluster does not have any regular shape and the individual atoms 
 exhibit largely diffusive motion. In addition, the cluster also
 shows a ``coexistence phase'', wherein after spending some time in 
 one of the phases the cluster spontaneously switches to the other
 and eventually back again.

   The ``phase changes'' in $\ar$ have been well characterized 
\cite{Berry1,efermi}. The three phases are best illustrated by the curve 
 of caloric equation of state- the plot of long time $(\sim 10^6 $ 
 time steps) average of kinetic energy against that of total 
 energy. The plot distinctly shows three regions- the solid-like
 region towards low kinetic/total energy, the liquid-like region
 on the higher end, and the coexistence phase in between. Another
 good diagnostic of the phase change is the root mean square bond
 length fluctuations, which shows a steep rise at the ``melting point'',
 indicating that there is a large increase in the mobility of the atoms
 of the cluster as it enters in to the liquid-like region.
 There are few other diagnostics given in detail in Ref. 
 \cite{Berry1} which demonstrate the occurance of the change of phase.
 
 The original motivation behind this work was to study the collective
 excitations in $\ar$ cluster as it passes through the phase changes.
 It was naively expected that in the transition region the system 
 would show a marked change in behavior, much like the bond length 
 fluctuations. However, this effort had to be abandoned as
 it was soon realized that the results obtained from different runs
 in the liquid-like region show a very large spread in values and hence
 it becomes hard to arrive at an unambiguous result. On the other hand, 
 we have had some surprising results in the solid-like region itself.
 We found that there is yet another, much lower characteristic temperature
 of the system, much more sharply demarcated on the total energy axis than
 the solid-liquid phase change. We shall argue that at this temperature
 the single particle excitations begin to destroy the collective
 excitations, and the effect is manifest as a qualitative change in 
 behavior of almost all the physical observables of the cluster.

  Details of the computational procedure are discussed in
  section II. The main results are presented in section III
  followed by a discussion. Finally we summarize the results and
  conclude.

\section{Computational Procedure}

 We perform isoergic molecular dynamics simulation of the 13-particle
 Argon cluster choosing the pairwise 6-12 Lennard-Jones interaction
\be{ V_{ij}(r_{ij})=4\epsilon \left[ {\sbr}^{12}-
 {\sbr}^6 \right]} \ee
with parameters $\sigma=3.4\times 10^{-8}$cm and $\epsilon=1.67
\times 10^{-14}$erg. The classical equations of motion were solved
using Verlet algorithm \cite{Verlet} with a time step of $2.0\times 
10^{-15}$sec. The total energy is found to be conserved to within 0.001\%.

 In an interacting many body system, we know that the average potential 
 of the static system contributes to the single particle excitations, and
 the interaction treated dynamically gives rise to collective
 excitations. Similarly it should be possible to think of the complex 
 dynamics of the cluster to be composed of collective motions in addition to
 the single particle undulations. Then we need to identify the
 collective modes of the cluster. In view of the nuclear deformations,
 it has been recognized \cite{Leder} that collective variables of an arbitrary 
 density distribution can be parametrised in terms of the moments of density
 distribution, and in particular for small deviations from spherical symmetry,
 expansion in terms of spherical harmonics components is the most natural
 description of the normal modes. Adapting the same idea, the contributions
 to different normal modes of the cluster as a function of time can be 
 projected out as
\be C_{lm}(t)=\int\rho({\bf r},t) Y^m_l(\theta,\phi) r^l dr \ee
where $\rho({\bf r},t)$ is the density distribution, which is
 discrete in the present case.

 Obviously the collective oscillations here are the shape oscillations.
 The monopole $(\ell =0)$ mode component is projected out simply as 
 the average radius of the cluster
 as a function of time, hence it corresponds to the uniform radial motion
 of the particles (``breathing'' mode). It should be noted at this
 point that in charged systems like the atomic nuclei or the metal
 clusters, the dipole mode $(\ell=1)$ usually is the strongest and the
 monopole is always very weak. In contrast the inert gas atom cluster
 $\ar$ has monopole mode as the strongest and the dipole mode simply
 does not exist. The reason is the following: A dipole motion in general
 corresponds to the vibration of the centroid of the distribution 
 about its mean position, and in case of charged systems, it is 
 the out of phase motion of the centroids of the opposite charge
 distributions that stabilizes the dipole oscillation. But
 there is no negative polarity for the mass distribution, so a dipole
 oscillation is ruled out, and so are all the modes with odd $\ell$,
 as they would correspond to the oscillation
 of the center of mass of the system about its origin.

 Here we present some observations made on the monopole oscillations of the 
 cluster. A monopole excitation is given to the cluster as follows:
 The system equilibrated at any temperature performs shape oscillations 
 of its own accord owing to the interplay of its kinetic and potential energies.
 First, a reference time t=0 is chosen such that at that time the cluster has
 expanded to its maximum, and is just about to start contracting. 
 At this stage, when most of its energy is in the potential energy form, 
 the cluster is given a sudden, radially uniform expansion, so as to raise
 the total potential energy of the cluster by a predetermined value.
 The velocities and hence the kinetic energies of the atoms are left
 unchanged, therefore the sudden expansion has the effect of 
 instantaneous increase of total energy by a predetermined value.
 The cluster responds to the increase in the amplitude and sets itself
 into oscillation in a pure mode and this 
 amounts to giving a monopole excitation to the cluster. We will add 
more on the method of giving the excitation to the system later in
the discussion.
 
 Fig. 1 shows the time evolution of the monopole component of the cluster
 at three different temperatures- the first one at $5^o$K,
 the second at $20^o$K, and the last at $30.5^o$K which is very 
 close to the melting point$(\sim 34^o K)$. 
 A monopole excitation corresponding to an excitation energy of
 $\delta E=0.05 \times 10^{-14}$ erg/atom is given at time t=0. The solid curve
 shows the time evolution of the cluster and one can immediately recognize
that the relaxation of the excitation resembles the time evolution
of a damped harmonic oscillator.
 The dotted curve in the plot actually is a fit to a damped oscillator
\be y(t)=y_0+A e^{-\lambda t} \cos(\omega't+\delta) \ee
 where $\lambda$ is the decay constant and $\omega'$ is the 
 reduced frequency of the damped oscillator. It should be mentioned
 here that the curves shown are actually not the ones obtained
 out of single runs, but in fact are the averages over 500 independent runs.
  Though the excitation is given at a configuration at which the 
 cluster is in a state of maximum expansion, in general not necessarily
 would all the constituents of the cluster be at their respective maximum
 displacements from the origin. As a consequence the response of the system
 to the excitation is quite sensitive to the initial configuration, and
 the quantities derived out of different runs show a 
 spread in value about their mean, increasingly so at higher temperatures.
 Hence it becomes necessary that some kind of ensemble averaging be done
 in order that good consistency is obtained. Note that the monopole 
 oscillation of the cluster is essentially simple harmonic at very
 low temperatures, and the oscillations are progressively damped
 as the temperature is increased. The fit is remarkably good,
 suggesting that one may map the monopole mode motion of the cluster to a
 1-dimensional under-damped harmonic oscillator. A plot of the potential energy
 of the cluster as a function of its radius as it evolves deviates very little
 from a perfect parabola [Fig. 2], giving further justification to the
 mapping. Hence we shall carry out further analysis in terms of the
 parameters of the equivalent harmonic oscillator, an exercise which would
 prove very fruitful. 
 
 From the fit to the time evolution of the excitation
 [Fig. 1] we already have the values of two of the parameters of the 
 equivalent harmonic oscillator, the damping coefficient $\lambda$ and the
 reduced frequency $\omega'$. From Fig. 2 we notice that the oscillator
 potential can well be taken to be parabolic with respect to the cluster
 radius, so we could make use
 of the Hooke's law to obtain the spring constant $k$ of the oscillator. We
 can go further. The reduced frequency $\omega'$ of a damped harmonic
 oscillator is related to its natural frequency $\omega$ by
 \be \omega'=\sqrt{\omega^2-\lambda^2} \ee
 and from the value of the natural frequency 
 we can obtain the mass $m$ of the equivalent oscillator from the
 relation $\omega=\sqrt{k/m}$. Thus, we now have the values of all the
 parameters of the equivalent oscillator, and henceforth we find it more
 convenient to discuss in terms of these parameters. We use the term
 ``reduced mass'' $m$ to refer to the mass of the equivalent oscillator 
 to distinguish it from the actual mass $M$ of the Argon atom.

  \section{Results and discussion}

  The results are summarized in Fig 3. The data are completely in
  the solid-like domain of the cluster (The last four points are from the 
  coexistence region). The abrupt change in qualitative
  behavior of the oscillator parameters at $E_{tot}=-5.444 \times 10^{-14}$
  erg/atom, which corresponds to a kinetic temperature of $T_s\sim 7.0
  ^oK$ is quite puzzling. The damping coefficient is zero for
  $T<T_s$ and then has a linear rise throughout the solid phase (Fig 3(a));
  the period of oscillation changes slope at $T_s$ (Fig 3(b)) ; the
  spring constant has a slow linear rise at low temperatures 
  but shows a dramatic $\sqrt{E_{tot}}$ rise beyond $T_s$ (Fig 3(c)) ;
  and finally, the reduced mass drops drastically at $T_s$ (Fig 3(d)).
  Some observations :
  \begin{enumerate}
  \item Of the four parameters of the damped harmonic oscillator, three
  (viz. the spring constant, the reduced mass and the damping coefficient)
  are completely independent of each other and only the frequency of
  oscillation is a derived quantity. However, the transition at $T_s$
  brings a qualitative change in all of them.
  \item It is on the total energy axis that these parameters indicate a
  clear-cut  transition. Therefore it is the total energy and not the kinetic
  temperature that is the most relavent parameter to study the system.
  \item The change of behavior along the total energy axis at this 
  transition is much too sharp compared to that  in the case of solid-liquid 
  phase change.
  \item Other dynamical properties like the rms bond-length fluctuations,
  velocity autocorrelation function, specific heat and Lyapunov
  characteristic exponent should also manifest
  the signature of this transition.  \end{enumerate}

 The sharpness of the transition at $T=T_s$ is unprecedented for a system 
 of its size. Earlier investigations do not seem to have paid much attention
 to the system at such low temperatures. From Fig 3(a) we find that at
 temperatures below $T_s$ the oscillations are completely undamped, which
 means that the collective mode is very stable and the individual atoms of
 the cluster perform periodic motion without losing coherence. However, it
 should be noted that anharmonicity is inherent in the underlying interaction
 potential, and manifests itself as a slow linear rise in the time period
 of oscillation with the increase in total energy.

 Then the question is what causes the damping of the collective mode at
 temperatures above $T_s$. Looking at the plot of the reduced mass,
 Fig 3(d), it is curious to note that below $T_s$ the mass of the 
 equivalent oscillator is smaller by a factor of 12 against the mass of the
 Argon atom, a factor same as the number of surface atoms. In addition, the
 value of the reduced mass drops drastically at $T_s$ and reaches an 
 asymptotic value of 1 at higher temperatures. This leads us to the picture
 of a set of a coupled oscillators, which continues to oscillate without
 losing strength if set-in in one of its normal modes; and the normal
 mode gradually decays if one or a few of the constituents are given
 additional independent disturbances. Based on this reasoning,
 we infer that for $T<T_s$ collective modes are stable whereas 
 above $T_s$, particles somehow begin to make independent motions and 
 these independent motions cause the damping of the collective mode.
 One can then argue that as the temperature is raised, these independent
 particle motions become more and more prominent, destroying the collective 
 oscillations faster than before. 

 If this argument is indeed correct, a Fourier analysis of motion of the
 particles should reflect this behavior. We took the power spectra of
 individual radial motion of the particles and plotted in Fig 4, 
 adding them together. The plot clearly shows 
 that for $T<T_s$ (Fig 4(a)), the spectrum 
 has a single sharp peak at the frequency component $\omega=11.3 \times 
 10^{10}$ Hz, and at $T_s$ (Fig 4(b)), additional components have just begun 
 to appear. At a slightly higher temperature (Fig 4(c)), the spectrum shows 
 a continuum at low frequencies, with a considerable reduction in
 the strength of the collective mode. The presence of an almost 
 flat continuous spectrum is a clear indication of the incoherent motion
 of the particles. However, it should be noted that
 there are also additional peaks in the spectrum, which could mean the
 presence of other collective modes. Indeed, most of the individual spectra
 do show prominence at these peaks. This may lead us to another scenario,
 that anharmonicity sets in at $T_s$; as a consequence of which other
 modes are excited due to mode-mode coupling, and thus causing the
 decay of the original pure mode. But it was found that
 the strengths are never the same at any component from the spectra 
 of any two of the particles. On the other hand the
 individual spectra of all the particles are stunningly identical at
 $T<T_s$. This gives a convincing evidence to the proposition that it
 is the onset of the independent particle motions that damps the collective
 oscillations at $T>T_s$. Finally, at still larger temperatures (Fig 4(d)),
 there is no trace of any collective mode.

 What remains to be explained then is how is it that these
 independent particle motions are triggered and
 why are they absent below $T_s$.
Interestingly, the situation is quite similar to some of the interacting
 quantum many body systems, like for example the case of a system with 
 magnetically ordered ground state, wherein the interaction gives rise to 
 a finite gap in the single particle excitation spectrum, besides 
 the formation of low lying collective modes. 

 Another surprising feature is the behavior of the spring constant above
 $T_s$. We know that the potential energy surface gets widened as the energy is
 increased, so we would expect the spring constant to decrease as the
 energy is increased. Instead, we notice that the spring constant increases,
 as square root of the total energy. To gain a better understanding, we plot
 the potential energy curves corresponding to different energies on the same 
 graph in Fig 5. Observe that for $T<T_s$, the oscillations at all 
 temperatures lie on the same potential energy curve, only the amplitude 
 increases with the increase in energy, just as in the case of an actual 
 harmonic oscillator. On the other hand, above $T_s$, the oscillations 
 of the cluster trace different potential energy curves at different 
 temperatures. This implies that it is not the same harmonic oscillator 
 any more at different temperatures. We reason that this 
 happens because the potential energy hyper-surface of the cluster in the
 3N dimensional co-ordinate space has a global minimum, and so long as the
 cluster is confined within this well the cluster sustains collective
 oscillations, individual particles retaining their relative phase coherence.
 Total energy $TE=5.444 \times 10^{-14} ergs/$atom, corresponding to $T_s$
 can now be seen as the depth of this global minimum well, within which
 pure normal mode of the cluster is stable.
 Above $T_s$, the cluster has sufficient energy to come out of this global
 minimum, and the particles find local minima pockets in the potential
 energy hyper-surface accessible to them. 
 Once they start making excursions to such local minima pockets, 
 the motion of different particles become asynchronous, causing them
 lose coherence, and this has the effect of damping the collective 
 oscillation.  These excursions are the ``single particle
 excitations'' of this classical system. 
 Different particles now go through the potential energy minimum at different 
 instances of time, and that is the reason why the potential energy curve is 
 much raised and the amplitude of the monopole mode is much smaller (Fig 5).
 However, it is not clear yet as to why, with the increase in total 
 energy,  the potential energy curve becomes narrower and 
 the spring constant increases as square root of total energy.
 
 The existence of a global minimum in the potential energy surface has long
 been known, the local minima are termed ``particle-hole structures'' 
 \cite{Berry2}. It has been accepted that the cluster has icosahedron
 structure so long as it is inside the global minimum and attains liquid-phase
 the moment it comes out of this well. Our results do not agree with this 
 view, in fact the depth of this global minimum is found to be just about
 $3.2\times10^{-14}~erg$, whereas it requires about $15.6\times10^{-14}~erg$
 of energy to take the cluster to the coexistence phase.

 At temperatures below $T_s$, since all the 12 surface atoms on the
 icosahedron execute coupled oscillations, the reduced mass $m$ of the 
 normal mode is given by 
 $${1\over m}=\sum_{i=1}^{12} {1\over M_i}={12\over M}$$
 or $${M\over m}=12$$ 
 in agreement with the results of the simulation (Fig 3(d)). The transition
 temperature $T_s$ corresponds to the situation wherein the motion of only
 one of the 12 surface atoms on the average becomes incoherent with the
 motion of the rest. This is reflected in the reduced mass, which at this
 temperature drops to one eleventh of the mass of the Ar atom, as can be
 seen from the figure. With the increase in temperature the reduced mass
 drops drastically and reaches an asymptotic value of 1. Since $(1/m)$ is
 a measure of the coherence of motion of the particles in the normal mode,
 the value of the ratio $(M/m)$ being 1.0 amounts to the constituents
 executing independent incoherent motion just as in the case of a viscous
 fluid. This transition to a liquid-like behavior at $T_s$, which is far
 below the melting temperature is a phenomenon new to the clusters, and is 
 absent in case of bulk solids. Hence it is expected that it will be more
 difficult to observe this phenomenon with the increase in cluster size.

 As has already been 
 mentioned we expect the transition at $T_s$ to manifest itself 
 in other observables as well. Here are the other results:
 \begin{itemize}
 \item The rms bond length fluctuation (not shown here) 
 starts from a value of zero and shows a quick rise in value till the 
 kinetic temperature $T_s$, and only then settles down to the slow 
 linear rise with temperature, as shown in Ref. \cite{Berry1}. 
 \item Fig 6 is the plot of velocity
 autocorrelation-- the solid curve just below $T_s$ and the dotted one just 
 above. This once again demonstrates the qualitative change in the
 dynamics of the cluster at $T_s$, there being a complete reversal of
 velocities at temperatures below $T_s$, and only a partial reversal
 the moment the temperature goes above $T_s$. A Fourier transform
 would show similar features as that of single particle radial
 coordinates.
 \item The average radius of the cluster has a linear temperature
 dependence all along the solid-phase; nevertheless shows a sudden 
 though small increase in the value at $T_s$.
 \item The maximal Lyapunov exponent (MLE) is close to zero at $T<T_s$
 and shows a sudden rise at $T_s$ (Fig 7).
 \end{itemize}

  It is quite satisfying that an independent dynamical quantity, MLE,
gives yet another illustration of the effect we have observed. It was
shown earlier that MLE data indicates solid-liquid phase change 
by a sharp jump in its value \cite{saroj}, and now we see that the
same dynamical quantity once again shows a steep rise-- starting
from a value of zero this time-- to mark the presence of another
transition. The MLE data implies that the system is integrable at $T<T_s$, and 
becomes chaotic above $T_s$. In other words, the onset of
single particle excitations drives the system into ergodicity.

Finally a few words about the method of giving excitation to the cluster.
It is essential that the cluster is set into a pure mode oscillation
to observe all the effects we have presented. In particular, if one
starts from a cluster equilibrated at $T>T_s$ and cools it below $T_s$,
it is unlikely that such a system would show the features as distinctly
as is presented here. The reason being that the single particle 
excitations present in the system above $T_s$ continue to be present
while the system is cooled and get proportionately amplified by the
radial displacement of the particles meant to provide the monopole
excitation. If started from low temperature side, this problem would
not arise until the temperature $T_s$, as the system continues to
perform monopole oscillations without losing strength. However, beyond
$T_s$, there is no way one could get rid of the presence of single
particle excitations to begin with, so we take recourse to the option
of taking statistical average over many independent runs, with the
hope that the effects of the independent particle motions are averaged
out, leaving behind the systematics. The essence of the analysis is 
the presence of a kinetic temperature $T_s$,
below which the cluster can sustain pure monopole oscillation for ever
and above which the particles suffer a loss of memory, invariably.

 \section{Summary}

In this article we have presented some results of the detailed analysis
of monopole excitation in $\ar$ cluster. Ensemble average of time evolution
of the monopole excitation is found to fit very well with a damped harmonic
oscillator. The parameters of the equivalent oscillator show a marked
change in behavior at kinetic temperature $T_s=7.0^oK$ which marks the
existence of yet another characteristic temperature in the system. Below
$T_s$ the cluster remains confined within the global minimum of the potential
energy surface, due to which the normal mode oscillations remain 
undamped. The velocity autocorrelation function shows a complete bounce-back
revealing that there is absolutely no loss of coherence. The system moves out
of the global minimum well at $T_s$, above which the local minima pockets in the
potential energy surface act as single particle excitation channels in
damping the collective modes. A continuous, flat power spectrum of the
radial motions of the particles at low frequency components
confirms the onset of these single particle excitations. The system is
naturally driven into ergodicity at this point due to the availability
of local energy minima pockets.

  This work paves way for many interesting questions. First of all,
one may like to investigate to see whether or not 
the presence of $T_s$ is a generic
feature of small clusters, irrespective of the interaction potential
used- including the {\it ab initio} dynamics. We obviously do not
expect this phenomenon to occur in large clusters. A closed shell structure
plays a vital role in this phenomenon, so we might still see it in 55
atom icosahedron cluster. One could also analyze higher angular
momentum components in the same fashion.

\vskip 1.5cm
{\bf Acknowledgments} We are grateful to V. Mehra and Prof. R.
Ramaswamy for providing us the MLE data on our request. One of us (UAS)
likes to thank Profs. S. D. Mahanti and D. G. Kanhere for some
stimulating discussions. We also thank the referee for his suggestions
to calculate the power spectra of the single particle coordinates and
the maximal Lyapunov exponent.

\newpage
{\bf Figure captions:}
\vskip 1cm
{\bf Figure [1].} Time evolution of the monopole oscillation at three different
temperatures: (a)at $5 ^oK$, (b)at $20 ^oK$ and (c)at $30.5 ^oK$. Average
radius of the cluster as a function of time simply projects out the time
evolution of the monopole mode. A monopole excitation is given at t=0.
Oscillations are simple harmonic and are undamped for 
$T<T_s(\approx 8.4^oK)$. 

{\bf Figure [2].} Potential energy of the cluster as a function of its
radius. The dotted curve is a perfect parabola. 

{\bf Figure [3].} Behavior of the parameters of the equivalent 1-d damped
harmonic oscillator as a function of total energy of the cluster. The data
are completely in the solid-like phase of the cluster (the last four points
are in the coexistence region). (a)Damping coefficient. There is no damping
for $T<T_s$ and a linear rise in damping coefficient with the total energy
for $T<T_s$. (b)Time period of oscillation. Shows a change of slope at 
$T=T_s$. (c)The ``spring constant'' $k$. It is a constant for $T<T_s$ but
shows a dramatic $\sqrt{E_{tot}}$ rise for $T>T_s$. (d)The ratio of the
mass of the Ar atom to the ``reduced mass'' of the oscillator. The value
of the ratio drops rapidly at $T=T_s$ and reaches an asymptotic value of
1.0 at high temperatures. The solid line is only a guide to the eye.

{\bf Figure [4].} Power spectrum of the time sequence of the radial
coordinates of the particles. Power spectra of the surface atoms are
added together and plotted. $P(\omega)$ is in arbitrary units and
total energy is in units of $1.0 \times 10^{-14} ergs/$atom. The 
onset of continous spectrum coincides with the damping of the
monopole excitation.

{\bf Figure [5].} Potential energy curves of the cluster at various 
temperaures $(^oK)$ : (a) 1.40 (b) 2.85 (c) 4.25 (d) 5.64 (e) 7.04 (f) 8.40
(g) 9.81 (h) 11.24 (i) 12.68 (j) 14.10 (k) 15.48. For
$T>T_s$, the cluster deviates away from the curve soon after one period.

{\bf Figure [6].} The normalised autocorrelation function. The solid curve is at
temperature just below $T_s$ and the dotted one just above $T_s$. The curves
are averages over 500 independent runs.

{\bf Figure [7].} The maximal Lyapunov exponent. The system is integrable
at temperatures below $T_s$ and chaos sets in at $T_s$. (Data from 
V. Mehra and R. Ramaswamy, private communications).

\end{document}